# Holographic indeterminacy and neutron stars

Scott Funkhouser
National Oceanic and Atmospheric Administration
2234 South Hobson Ave.
Charleston, SC, 29405-2413

ABSTRACT
The holographic indeterminacy resulting from the quantization of spacetime leads to an inherent uncertainty $(l_p L)^{1/2}$ in the relative positions of two events, separated by a distance $L$, in a direction transverse to a null ray connecting the events, where $l_P$ is the Planck length. The new indeterminacy principle leads to a critical condition in which the holographic uncertainty in the relative transverse positions of two diametrically opposed particles on the surface a body becomes greater than the average distance between particles in the body. The Chandrasekhar mass and the characteristic nuclear density emerge as the minimum mass and density of a baryonic body that could meet the critical criteria. Neutron stars are therefore identified as a class of bodies in which holographic indeterminacy may have physical consequences.

*1. A unified uncertainty relation*

The quantization of spacetime may impose a new uncertainty principle that is fundamentally holographic in nature [1]. Consider two events separated by some distance $L$ that is very large with respect to the Planck scale, and let **y** represent some direction that is transverse with respect to a null trajectory connecting the two events. Considerations of a basic model of the structure of an emergent, quantum mechanical spacetime introduce indeterminacy in the relative transverse orientation a null rays connecting the two events [1]. Consequently, the relative transverse position $y$ of two events separated by a distance $L$ is indeterminate, and the associated uncertainty is characterized by a standard deviation $\Delta y$ that is limited by [1]

$$\Delta y > (l_P L)^{1/2} \qquad (1)$$

where $l_P$ is the Planck length. The indeterminacy leading to (1) should generate a pervasive "holographic noise" that is manifest as ubiquitous shear perturbations characterized by a uniform spectral power-density $l_P/c$, where $c$ is the vacuum-speed of light [1].

The purpose of this present work is to investigate additional possible effects of the indeterminacy in (1). Consider some region of total mass $M$ and radius $R$ consisting of particles whose average mass is $m$. The average number-density of particles within the region is $n=\rho/m$, where $\rho=3M/(4\pi R^3)$. Let there be some arbitrary test particle, either inside or outside of the spherical region, that is located a distance $L$ from an arbitrary particle within the region at some moment. Let $y$ represent the relative transverse position of the two particles. There exists a critical distance $L=L_h$ at which the uncertainty $(l_P L)^{1/2}$ in the transverse position of the particle relative to the reference event is equal to the average distance $d=(4\pi n/3)^{-1/3}$ between particles. The critical distance $L_h$ is given by

$$L_h = \frac{1}{l_P n^{2/3}}. \qquad (2)$$

Consideration of $L_h$ in relation to the size of the region leads to a critical condition $L_h=2R$ that may have physical consequences. If $L_h$ is only slightly smaller than $2R$ then the uncertainty in the relative transverse positions of any two diametrically opposed particles on the surface of the region would be greater than $d$. If $L_h$ is smaller than $2R$ by more than just a microscopic amount then the relative transverse uncertainties of a significant number of distinct pairs of particles within the body would be greater than $d$.

Suppose that there exists some physical body for which $L_h<2R$. Consider some arbitrary constituent particle $A$ that is separated from another constituent particle $B$ by a distance greater than the critical length $L_n$. Relative to $A$, the transverse coordinates of the particle $B$ and the particles in the immediate vicinity of $B$ must be, on average, overlapping, and *vice versa*. That situation may lead to a problematic inconsistency for the structure of the body if there should exist a large number of distinct pairs $A$ and $B$. The specific physical circumstances under which such a situation could occur are explored in the following paragraphs, and the issue of the overlapping coordinates is found to be prohibitive.

It follows from (2) that the condition $2R>L_h$ is satisfied if

$$R > \frac{1}{2}\left(\frac{3}{4\pi}\right)^{2/3}\frac{m^{2/3}}{l_P \rho^{2/3}} \equiv R_h(\rho,m), \tag{3}$$

where $R_h(\rho,m)$ represents a lower bound on the radius of the body that could exhibit the critical holographic behavior. The criterion in (3) may be expressed equivalently in terms of the associated mass $M=4\pi\rho R^3/3$ as

$$M > \left(\frac{3}{32\pi}\right)\frac{m^2}{l_P^3 \rho} \equiv M_h(\rho,m). \tag{4}$$

For $m$ equal to the nucleon mass $m_n$, the lower bounds $M_h(\rho,m)$ and $R_h(\rho,m)$ are enormous and unphysical except for very large densities $\rho$. Any body that could possibly approach the point near which $2R>L_h$ must be therefore extremely dense.

In order for any physical body to satisfy (3) and (4), the ratio $M/R$ must be less than $c^2/(2G)$, where $c$ is the vacuum speed of light and $G$ is the Newtonian gravitational coupling. It follows from $M=4\pi\rho R^3/3$ that the requirement $M/R<c^2/(2G)$ leads to an upper bound on $R$,

$$R < \left(\frac{3c^2}{8\pi G\rho}\right)^{1/2}. \tag{5}$$

(An upper bound on $M$ follows also from (5), but it is not necessary for the present considerations.) In order $R$ to satisfy both (3) and (5), the right side of (5) must be larger than $R_h(\rho,m)$, which is satisfied only for

$$\rho > \left(\frac{3}{32\pi}\right)\frac{m^4 c^3}{\hbar^3} \equiv \rho_h(m), \tag{6}$$

where $\hbar$ is the Planck constant. The lower bound $\rho_h(m)$ represents the smallest average density of a physical body that could exhibit the critical holographic behavior $2R>L_h$. It follows from (6) that the lower bounds $R_h(\rho,m)$ and $M_h(\rho,m)$ must be limited respectively according to

$$R_h(\rho,m) < \frac{c^2}{2G}\frac{m_P^3}{m^2} \equiv R'_h(m) \tag{7}$$

and

$$M_h(\rho,m) < \frac{m_P^3}{m^2} \equiv M'_h(m). \tag{8}$$

If the average density $\rho$ of a body is greater than the lower bound $\rho_h(m)$ and the radius $R$ of the body is greater than $R_h(\rho,m)$ then the associated critical length $L_h$ would be smaller than $2R$. (Note that $R>R_h(\rho,m)$ is equivalent to $M>M_h(\rho,m)$.) The requirement $\rho>\rho_h(m)$ is nearly prohibitive since $\rho_h(m)$ characterizes roughly the greatest possible scale of density that could be achieved by any physical body. (Note that $\hbar/(mc)$ is proportional to the Compton wavelength $l_m$ of the particle, and the minimum density $\rho_h(m)$ is therefore proportional to $m/l_m^3$.) There is, however, one class of astronomical bodies whose parameters approach the critical criteria. For $m$ equal to the nucleon mass $m_n$ the minimum critical density $\rho_h(m)$ is approximately $5.2\times10^{18}$kg/m$^3$, which corresponds roughly to the maximum density expected in the core of a neutron star. In order for a body whose average density is just less than $\rho_h(m)$ to exhibit the critical holographic behavior, the mass of the body must be greater than $M_h'(m_n)$, which is approximately $3.7\times10^{31}$kg. That mass is of order near the characteristic stellar mass, and it is plausible that some neutron star could have a mass near $M_h'(m_n)$. Note also that

$$M_h'(m_n) = \frac{m_P^3}{m_n^2} \tag{9}$$

is proportional to the Chandrasekhar mass limit.

If there should exist a neutron star that satisfies the critical conditions outlined in the preceding paragraph then the relative transverse positions of a certain number of distinct pairs of particles within the body must overlap with their respective neighboring particles. That situation would be problematic since neutron stars are, in general, degenerate, and the structure of a degenerate body is very sensitive to the spacing between its constituent particles. Furthermore the constituent particles of a degenerate body are, by definition, arranged as closely as the basic principles of quantum mechanics allow. Even though the overlapping associated with holographic indeterminacy is a non-local, relative effect that is manifest at separations greater than $L_h$, it may still have consequences since it would affect the internal consistency of the structure of the body. It is important to note that neutron stars are relativistic bodies, and a typical neutron star may feature various levels of internal structure, including perhaps an exotic core containing superconducting protons and a quark plasma. It is therefore not reliable to draw more specific conclusions without a more detailed treatment. The present analysis is, however, useful in demonstrating that there is at least one astrophysical regime in which holographic indeterminacy may be important. It is also plausible that the present considerations establish roughly an upper bound on the density and mass of a neutron star. If holographic indeterminacy should result in a problematic overlapping of transverse coordinates then it may be that nature prevents inherently a neutron star from satisfying the criteria associated with the critical condition $2R>L_h$.

___________________